\RequirePackage{fix-cm}
\documentclass{svjour3}                     
\smartqed  
\usepackage{graphicx}
\usepackage{amssymb}
\usepackage{amsmath}
\usepackage[numbers]{natbib}
\usepackage{textcomp}
\usepackage{latexsym}
\journalname{Journal}

\usepackage{listings}
\lstnewenvironment{Code}[1][]{
\lstset {
     float=htp,
     frame=single,
     aboveskip=-5pt,
     belowskip=-5pt,
     framextopmargin=1pt,
     framexbottommargin=1pt,
     basicstyle=\scriptsize\ttfamily,
     captionpos=b,
     breaklines=true,
     language=c++,
     morekeywords={*,pragma, redthreads, rolex},
     #1
  }
} {}

\lstnewenvironment{CodeExample}[1][]{
\lstset{
     float=h,
     frame=single,
     basicstyle=\scriptsize\ttfamily,
     captionpos=b,
     breaklines=true,
     commentstyle=\textit,
     framextopmargin=1pt,
     framexbottommargin=1pt,
         morekeywords={*,resilience, redthreads, robust, share, private, compare},
         belowskip=-0.8 \baselineskip,
     #1
  }
} {}

\begin{document}

\title{RedThreads: An Interface for Application-level Fault Detection/Correction through Adaptive Redundant Multithreading
\thanks{The authors would like to acknowledge the support for this work provided through Scientific Discovery through Advanced Computing (SciDAC) program funded by U.S. Department of Energy, Office of Science, Advanced Scientific Computing Research under award number DE-SC0006844. Partial support for this work was also provided by the US Army Research Office (Award W911NF-13-1-0219) \\ \\
Sandia National Laboratories is a multi-program laboratory managed and operated by Sandia Corporation, a wholly owned subsidiary of Lockheed Martin Corporation, for the U.S. Department of Energy's National Nuclear Security Administration under contract DE-AC04-94AL85000.}}
\titlerunning{RedThreads: An API for Adaptive Redundant Multithreading}   

\author{Saurabh Hukerikar* \and
        Keita Teranishi$^{\dagger}$  \and
	Pedro C. Diniz* \and
        Robert F. Lucas*
}

\institute{*Information Sciences Institute \at
           University of Southern California \\
           4676 Admiralty Way Suite 1001, \\
           Marina del Rey, CA 90292 USA \\
           \email{saurabh, pedro, rflucas@isi.edu}           
           \and
           $^{\dagger}$Sandia National Laboratories \at
	   7011 East Avenue,      \\
           Livermore, CA 94551 USA \\
	   \email{knteran@sandia.gov} 
}

\date{Received: date / Accepted: date}

\maketitle

\begin{abstract}
In the presence of accelerated fault rates, which are projected to be the norm on future exascale systems, it will become increasingly difficult for high-performance computing (HPC) applications to accomplish useful computation. Due to the fault-oblivious nature of current HPC programming paradigms and execution environments, HPC applications are insufficiently equipped to deal with errors. We believe that HPC applications should be enabled with capabilities to actively search for and correct errors in their computations. The redundant multithreading (RMT) approach offers lightweight replicated execution streams of program instructions within the context of a single application process. However, the use of complete redundancy incurs significant overhead to the application performance. 

In this paper we present RedThreads, an interface that provides application-level fault detection and correction based on RMT, but applies the thread-level redundancy adaptively. We describe the RedThreads syntax and semantics, and the supporting compiler infrastructure and runtime system. Our approach enables application programmers to scope the extent of redundant computation. Additionally, the runtime system permits the use of RMT to be dynamically enabled, or disabled, based on the resiliency needs of the application and the state of the system. Our experimental results demonstrate how adaptive RMT exploits programmer insight and runtime inference to dynamically navigate the trade-off space between an application's resilience coverage and the associated performance overhead of redundant computation.

\end{abstract}

\section{Introduction}

The demand for higher computational power to enable larger and more accurate simulations, visualizations and analyses in a variety of domains of science and engineering is driving the high-performance computing (HPC) community towards exascale capability supercomputers \cite{ASASC:2010} \cite{Dongarra:2011:IES}. Existing petascale-class HPC systems already employ millions of processor cores and memory chips to drive HPC application performance. Recent architectural trends suggest that future exascale HPC systems will be built from hundreds of millions of components organized in complex hierarchies. With an exponential increase in the number of components, the overall reliability of the system decreases proportionally. Furthermore, with process technology scaling, these systems will be constructed from VLSI devices which are less reliable than those used today \cite{Borkar:2005:Micro}. Long-running scientific applications on these large scale systems are therefore increasingly likely to incur errors during their execution, threatening the validity of the scientific simulations and limiting application scalability \cite{DARPA_ExascaleTechStudyReport:2008}. The occurrence of frequent faults will cause errors in the applications' program state, which will cause their executions to crash, or worse, to run to completion with incorrect results.

HPC programming paradigms and execution environments are designed with the expectation of correct behavior from the underlying system layers. Due to their fault oblivious designs, HPC applications lack sufficient mechanisms to detect errors and limit their propagation. In the presence of extremely high fault rates, which are anticipated in future exascale-class HPC systems \cite{DARPA_ExascaleResilienceStudyReport:2009}, the most commonly used approaches such as checkpoint and rollback recovery (C/R) and algorithm-based fault tolerance (ABFT) will be insufficient, or in certain cases incapable in dealing with all the faults and errors in the system. We previously proposed a complementary approach, called Rolex \cite{Hukerikar:Rolex:2016}, which provides a set of resilience-oriented language extensions. Through these language extensions, which include type qualifiers, compiler and execution directives, library routines and environment variables, we supported various fault management semantics. In this paper we define and evaluate RedThreads, a specification for an API for redundant multithreading (RMT) that permits in-situ application-level error detection and correction in C/C++ programs. The RedThreads API is syntactically derived from Rolex \cite{Hukerikar:Rolex:2016} and includes language-level features that offer lightweight redundant execution streams of the program's code regions. RedThreads supports thread-level dual and triple-modular redundancy (DMR and TMR). The programming constructs provide a succinct API to specify regions of code for which error detection/correction is necessary, shielding HPC programmers from the low-level threading interfaces. The constructs are supported by a compiler infrastructure and runtime system, which through the adaptive use of RMT deliver flexibility and scalability for HPC application codes. The adaptive RMT capabilities allow dynamic tuning of the fault detection/correction coverage of the application, and consequently the ability to trade-off performance overhead due to redundant computation with the application robustness. 

The remainder of this paper is organized as follows: Section \ref{sec:Redundancy_Basics} describes the concepts and terminology used in redundant computing. Section \ref{sec:RedThreads_API} describes the API, including the syntactic structure of the directives, their semantics, as well as provides example codes in which the error detection/correction are embedded using the RedThreads API. Section \ref{sec:CompilerSupport-RMT} describes the compiler-based construction of the structured code blocks executed using adaptive RMT. Section \ref{sec:Runtime} details the design of the runtime system, including the supporting library routines and the various RMT scheduling policies and Section \ref{sec:Optimization} describes runtime-based optimization strategies. Section \ref{sec:Experimental_Evaluation} presents the experimental infrastructure, evaluation methodology and the performance results. Section \ref{sec:Related_Work} surveys related redundant multithreading approaches in the context of HPC applications and systems.

\section{Background}
\label{sec:Redundancy_Basics}

\subsection{The Redundancy Solution for Fault Detection/Correction}
Redundancy is a well-studied canonical solution \cite{vonNeumann:1956}, in which the computation is performed by three redundant modules and called triple modular redundancy (TMR). The more general form of this approach is called n-modular redundancy (nMR). Redundancy is a compensation strategy through which errors in the system state may be detected (when there is a mismatch in the outputs of the modules) and corrected (by performing majority voting on the results from each module). These approaches are widely used in critical application domains, such as space missions, aviation, medical control systems, in which the result of failure is usually catastrophic, and the need for correctness is independent of the likelihood of failure. 

In the world of HPC, studies have demonstrated the use of multi-modular redundancy in compute nodes accommodates a reduction in individual component reliability by a factor of 100-100,000, but incurs 2x or 3x increase in performance costs and energy \cite{Engelmann:2009}. Partial process replication was also explored, but these studies have noted that there is no alternative to complete process replication for highly resilient operation \cite{Ferreira:2011,Stearley:2012}. Therefore, beyond exploratory studies, macroscale compute node-level or process-level multi-modular redundancy approaches have not been widely used in production HPC system environments, due to the significant overhead to performance and energy incurred by the replication of every MPI rank of an application. 

\subsection{Redundant Multithreading}
Redundant Multithreading (RMT) uses multiple thread contexts for fault detection and correction. Identical copies of partial or complete program code are executed using separate thread contexts. For error detection, the outputs values produced by duplicate threads are compared and upon a mismatch the checker flags an error. This notification can be used to initiate a hardware or software-based recovery sequence. The TMR equivalent of RMT entails execution of the program code by three independent thread contexts. By selecting the result through majority voting, incorrect outputs produced by either thread copies may be filtered out.  

\begin{figure} [tp]
\centering
\includegraphics[width=50mm,height=45mm]{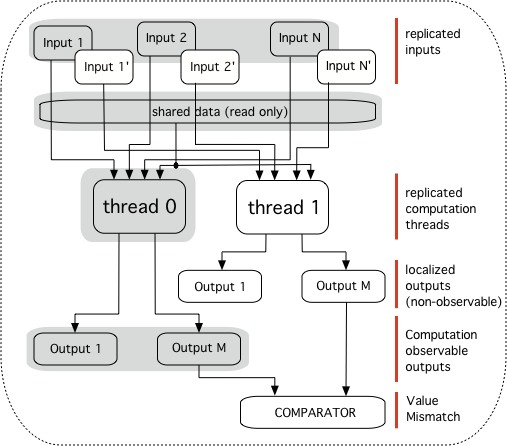}
\caption{Sphere of replication}
\label{Fig:Sphere_of_Replication}
\end{figure}
One of the key concepts in RMT is the \textit{sphere of replication} \cite{Mukherjee:2002}, which represents a logical boundary that includes a unit of computation that may be replicated (Figure \ref{Fig:Sphere_of_Replication}). Any fault that occurs within the sphere of replication propagates to its boundary. The computation enclosed within the sphere of replication is executed by the redundant threads. The inputs may be replicated, or a single copy may be shared between the redundant threads. Errors are detected by comparing specific outputs produced by the redundant execution of the spheres of replication. Any faults that do not manifest themselves at the boundary of the sphere in the output values get masked and are treated as benign in terms of their impact on the application outcome. The replication of the inputs and the computation remain transparent to the program state outside the sphere of replication. The process of managing the size and extent of the sphere of replication is important because it affects the overhead to application performance and extent of fault coverage offered. The design considerations are:
\begin{enumerate}
\item \textit{For which parts of the application will the redundant execution mechanism detect faults?} \\
Based on the fault coverage requirements of the users and algorithmic features of the code, the sphere of replication may include the complete program execution, or only the code for a specific function body, or a basic block of instructions. The extent of redundant computation proportionally increases the overhead to application performance and energy. 

\item \textit{What are the inputs to the sphere of replication, and do they need to be replicated?} \\ 
The selection of inputs has important implications for the application performance since it increases the number of memory accesses by the program. On the other hand, the failure to replicate any inputs may potentially lead the redundant threads on divergent execution paths. Therefore, inputs need to be selectively replicated, or they should be protected using other redundancy mechanisms such as checksums, parity, etc.

\item \textit{What are the output values from the sphere of replication that need to be compared in order to detect the presence of faults in the computation?} \\ 
In general, any values that are computed within the sphere of replication must be compared in order to check for a mismatch. The failure to compare meaningful application values potentially compromises fault coverage and the usefulness of the redundant execution. However, unnecessary comparisons increase the overhead to the application performance without improving fault coverage.
\end{enumerate}

Hardware implementations of RMT leverage simultaneous multithreading (SMT) capabilities \cite{Vijaykumar:2002} in which the processor architecture is designed to fetch instructions from multiple thread contexts, and their execution is interleaved. The threads may even be mapped to independent cores in the context of a chip multiprocessor (CMP), called chip-level redundant multithreading (CRT) \cite{Reinhardt:2000}. Hardware-based RMT is transparent to the operating system and the application, but incurs significant overhead due to instructions flowing through complex processor pipelines, and due to the increased contention for shared resources such as caches. Software-based approaches use threading library interfaces to create redundant threads within a single OS process. The redundant threads are loosely coupled since the intermediate thread state need not be synchronized at every clock cycle, and therefore offer flexibility in terms of scheduling the redundant threads. The partial sharing of state between the redundant threads also helps in mitigating some of the runtime overhead by reducing additional memory accesses. However, even with software-based RMT, the complete replication of the entire program code still incurs significant overhead to the overall application execution, particularly in long-running HPC applications executing on large-scale systems. The significant barrier to the adoption of RMT in HPC is the lack of convenient mechanisms that allow tuning its use and the scope of the sphere of replication to dynamically trade-off application reliability and performance based on runtime information or user-defined policies.

\section{RedThreads: An API for Adaptive Redundant Multithreading}
\label{sec:RedThreads_API}

\subsection{Overview: Programmer Managed Scoping of Redundancy}
\begin{figure} [tp]
\centering
\includegraphics[width=80mm,height=60mm]{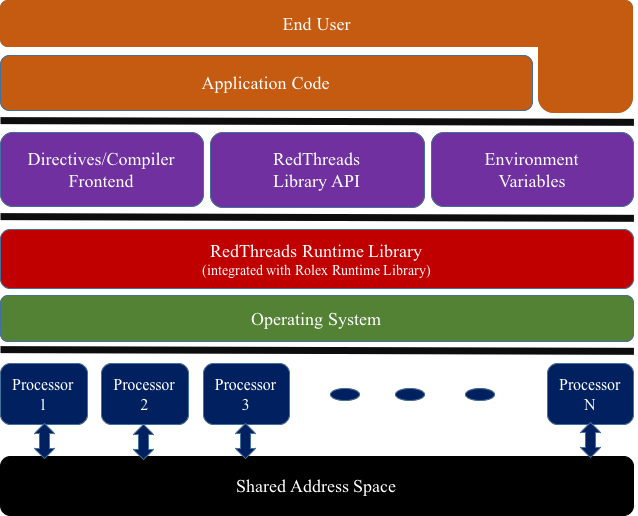}
\caption{The Adaptive Redundant Multithreading Solution Stack}
\label{Fig:RedThreads_Solution_Stack}
\end{figure}
Hardware-based RMT as well as automatic compiler-based approaches that introduce redundancy into programs inadequately understand the algorithm structure and therefore such methods tend to be insufficient in finding the balance between effective fault coverage and performance overhead. The goal of RedThreads is to provide a simple API to capture fault detection/correction intent in HPC programs. The API seeks to provide mechanisms that enable scientific programmers and library developers to leverage their domain knowledge and understanding of their codes to selectively apply redundant computation on specific routines as well as choose which input variables warrant replication and identify output variables that must be compared to infer the presence of errors in the program execution. 

Our work on Rolex \cite{Hukerikar:Rolex:2016} laid the foundation for explicit resilience-oriented programming models. The RedThreads API extends this approach, through directives, library routines and environment variables, to incorporate RMT in C and C++ programs. This application-driven approach enables the resilience to be tuned to the needs of the applications and the users. The solution stack (Figure \ref{Fig:RedThreads_Solution_Stack}) is based on the combination of language-level directives and library routines together with an intelligent runtime system and a compiler infrastructure. This adaptive RMT solution provides HPC applications with the capability to tune the use of redundancy for error detection and correction opportunistically to mitigate the overheads to application performance.

In designing the language extensions, we sought to embed the RMT constructs within higher level programming models rather than expose the programmer to raw threading interfaces. The API is designed to greatly simplify incorporating error detection/correction based on redundancy by enabling the programmer to focus on the resilience of an algorithm and the need for error detection/correction rather than the specific mechanisms for thread creation, synchronization and destruction. Additionally, the emphasis on high-level abstractions enables portability to alternative low-level threading libraries, compilers and operating systems, and to any hardware architecture. With the RedThreads directives and library routines, the RMT may be embedded within existing programming model features which minimizes changes to program code structure. This promotes predictability of good performance and scalability of the performance to larger systems. We were also minded to ensure composability with other HPC programming paradigms such as OpenMP and MPI as well as productivity libraries such as BLAS and LAPACK.

\subsection{Syntactical Structure of RedThreads}
The RedThreads API consists of compiler directives and library routines. The API is specified for C/C++ since these are the languages in which many new mainstream performance-oriented applications are written. We also support bindings for FORTRAN. The syntax of the RedThreads directives to specify spheres of replication is as follows:\\ 
\begin{CodeExample}[label = {lst:RMT_Syntax},frame=single]
#pragma redthreads detect|correct share( variable_list ) private( variable_list ) compare ( variable_list )
{   /* statement list */  }

#pragma redthreads declare detect|correct 
  /* C/C++ function definition or declaration */
\end{CodeExample}

The directives begin with the {\tt redthreads} keyword to indicate the use of adaptive RMT. The directives follow the conventions of the C and C++ standards for compiler directives. They contain a {\tt STRENGTH} clause to specify whether a detection or correction capability is required, which implicitly indicates the number of redundant thread copies that the runtime must create. The directive format provides the application programmer with the flexibility to control every aspect of the sphere of replication, including the scope of computation, the input and output variables. Therefore, the directives include data scoping clauses that are used to impose rules on whether the data variables are replicated among the redundant executions of the structured block, as well as to identify output variables produced by the sphere of replication, which need to be compared in order to detect the presence of errors. The scope of the sphere of replication applies to at most one succeeding statement following the directive, which must be a structured block that is enclosed in a pair of `\{' and `\}'. The \textit{structured block} is defined as a C/C++ executable statement which may be a compound statement, but has a single point of entry at the top and a single exit point at the bottom. The point of entry cannot be the target of a branch and no branch is allowed from within the block, except for program exit. Instances of the structured block can be compound statements including iteration statements, selection statements, or try blocks. The declarative directives may be associated with function declarations and definitions. These also enable the compiler to create multiple versions of the specified C/C++ function, one of which includes RMT execution. 

\subsection{Semantics}
The RedThreads directives enable the HPC application programmers to selectively apply redundancy on specific code regions to detect/correct errors in the computation of the corresponding application phases. When the compiler encounters a RedThreads directive, it creates explicitly redundant multithreaded code. The code statements contained in the structured block are optionally executed by redundant threads. 

\begin{figure} [tp]
\centering
\includegraphics[height=30mm,width=\linewidth]{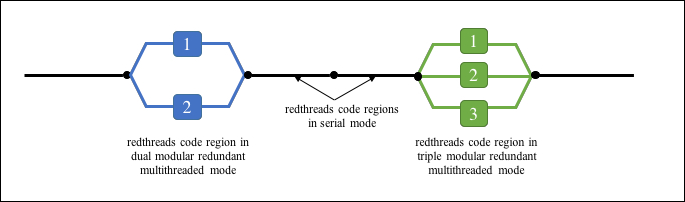}
\caption{Adaptive Redundant Multithreading Semantics based on Fork-Join Threading Model}
\label{fig:RedThreads_Semantics}
\end{figure}

RedThreads is based on the shared memory machine model, in which all parts of application have access to the same memory address space. RedThreads uses the fork-join model of parallel execution in which the program begins execution as a single execution stream until it encounters a {\tt redthreads} region. The statements in the program that are enclosed by the {\tt redthreads} region construct are then executed in parallel by independent threads. There is an implied barrier at the end of a RedThreads region to synchronize the threads and compare values produced by each thread copy. This is illustrated in Figure \ref{fig:RedThreads_Semantics}. The parallel redundant execution of the application region is transparent to user. Additionally, the API permits the runtime environment to dynamically alter the number of threads used to execute the code regions. This enables thread parallelism to be added incrementally during application execution for the redundancy to be strengthened, or even be disabled. 

The RedThreads API enables two models of reliability that are relevant to HPC applications:
\begin{itemize}
\item \textbf{In-situ detection/correction}, which enables the application to transparently detect, and potentially correct, errors in the user-visible program state in a very targeted fashion by scoping spheres of replication to match critical application phases. The RedThreads directives enable the incorporation of redundancy-based error detection/correction capabilities that are free of any distortions that may otherwise be necessary to map algorithms to threading interfaces. 

\item \textbf{Selectively reliable computation}, in which the computation may be partitioned into \textit{reliable} and \textit{unreliable} phases. For example, FT-GMRES \cite{Hoemmen:2011} allows the inner solver step to return an incorrect solution caused by error events, with the expectation of reliable execution from the outer iteration. The reliable phases may be executed in redundant multithreaded mode to enable the application to detect invalid values in the application's variable state. 
\end{itemize}

\subsection{Examples}
\subsubsection{Double Precision General Matrix\allowbreak-\allowbreak Matrix\allowbreak Multiplication\allowbreak (DGEMM)\allowbreak}
\begin{figure}[h]
\centering
\includegraphics[width=\linewidth,height=30mm]{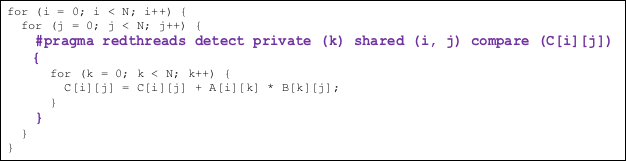}
\caption{Application of RMT directive for DGEMM}
\label{Fig:DGEMM_Code}
\end{figure}

In a na\"{i}ve implementation of the basic matrix-matrix multiplication, which consists of three loop levels, the innermost loop steps through one row of matrix A and one column of matrix B over a loop variable {\tt k}. The inner loop calculates a dot product of the row of A and the column of B and generates one element of result matrix C. We define the redthreads pragma block to include this inner dot product, \textit{i.e.,} the computation resulting from the multiplication of a single row and single column of the operand matrices (as illustrated in the code snippet in Figure \ref{Fig:DGEMM_Code}).

\subsubsection{Sparse Matrix Vector Multiplication (SpMV)}
\begin{figure}[h]
\centering
\includegraphics[width=\linewidth]{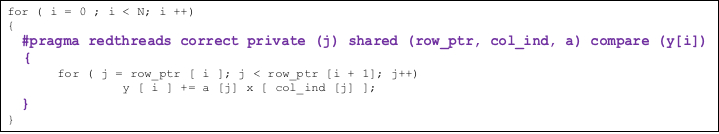}
\caption{Application of RMT directive for SpMV}
\label{Fig:SpMV_Code}
\end{figure}

In the implementation of SpMV shown in Figure \ref{Fig:SpMV_Code}, the SpMV the computation iteratively multiplies a constant sparse matrix with an input vector. Assuming that the sparse matrix is represented in the compressed storage row (CSR) format, we place the body of the nested for-loop, i.e., the inner product, in the {\tt \#pragma redthreads} structured code block. Given the memory-bound nature of SpMV, we only include the loop counter variables in the {\tt private} clause while the remaining data variables are shared between the redundant threads. The reduced vector element y[i] is included in the {\tt compare} clause in order to check the outcome of every iteration of the outer loop. 

\section{Compiler Support for Adaptive Redundant Multithreading}
\label{sec:CompilerSupport-RMT}

\begin{figure} [tp]
\centering
\includegraphics[width=\linewidth,height=30mm]{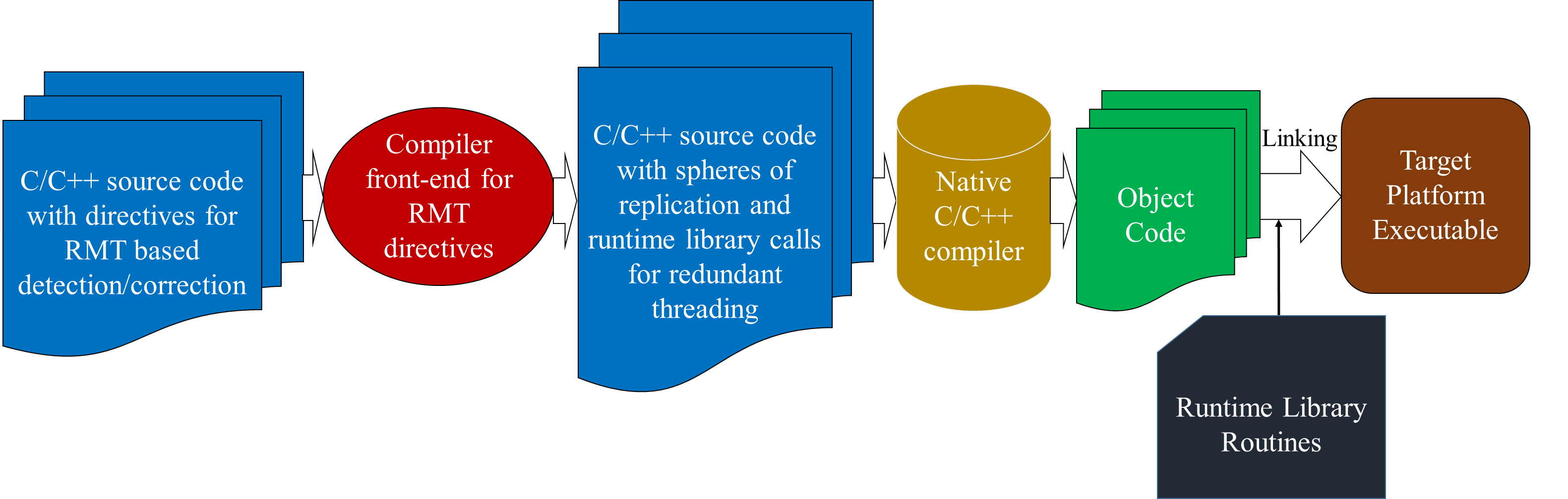}
\caption{Compiler Infrastructure for Redundant Multithreading}
\label{Fig:Compiler-Arch}
\end{figure}

\subsection{Construction of Application Programmer Scoped Spheres of Replication}
The compiler plays the role of a key intermediary that propagates the knowledge expressed by the programmer in creating spheres of replication to the runtime system. While the creation of spheres of replication through the language-level pragma directives permits HPC programmers to make strategic decisions on their scope, it is the compiler that processes the directive and uses it to create explicitly redundant multithreaded code with error checking. The compiler shields the details of the creation of the RMT from the application developer. The compiler infrastructure is responsible for the redundant thread creation, execution and termination. It also manages the shared, private data variables and inserts code statements that compare the data variables listed in the {\tt compare} clause. However, the compiler does not check for synchronization or correctness.

We have extended the compiler front-end for Rolex to support redundant multithreaded execution of the spheres of replication through source-to-source code transformations. The overview of the compilation process is illustrated in Figure \ref{Fig:Compiler-Arch}. The front-end generated code only uses base language constructs and calls to runtime library routines so that a native C/C++ compiler can be used to generate code for the target platform. Using a two-stage compilation process enables programs to be compiled using a standard C/C++ compiler infrastructure and leverages various implementations of threading libraries and enables portability to various architectures. The front-end source-to-source translators are built using the ROSE compiler infrastructure \cite{ROSE:Compiler}.

\begin{figure} [tp]
\centering
\includegraphics[width=\linewidth,height=65mm]{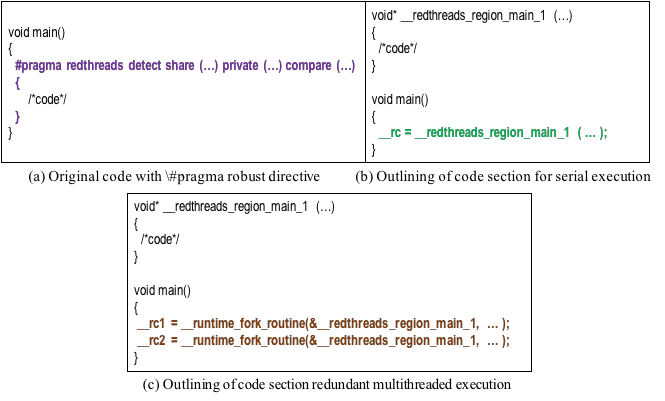}
\caption{Code Outlining for Redundant Multithreaded Execution}
\label{Fig:Code-Outlining}
\end{figure}

In order to implement the sphere of replication from the structured block specified after the directive, the compiler \textit{outlines} the code in structured block. Outlining entails extraction of the code segment, i.e., the statement list from a host function, and creation of a new function referred to as the outlined function. The original code segment in the host function is typically replaced with a call to the outlined function \cite{Liao:2010}. Figure \ref{Fig:Code-Outlining}(a) and (b) show the outlining transformation. To support RMT, the original code segment in the host function needs to be replaced by a call to a threading library routine to which the outlined function pointer is passed as argument. This code transformation is shown in Figure \ref{Fig:Code-Outlining}(c). Each redundant thread executes the outlined function, and the data variables specified in the \texttt{share} clause are passed as arguments to this procedure. The data variables specified in the \texttt{private} clause are stored on the thread's stack. Although outlining introduces some overhead to the application execution, it makes the translation of the structured block into redundant multithreaded code and the management of the data scoping straightforward.

\subsection{Support for Adaptive RMT Execution of Spheres of Replication}  
\begin{figure} [tp]
\centering
\includegraphics[height=75mm]{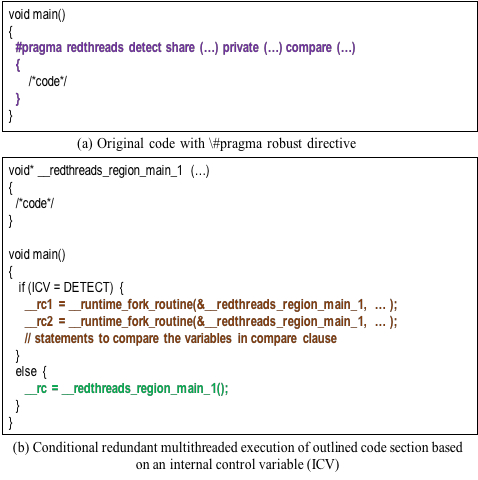}
\caption{Code Outlining and Transformations for Adaptive Redundant Multithreaded Execution}
\label{Fig:Code-Outlining-Adaptive}
\end{figure}

In addition to outlining the structured block into spheres of replication, the compiler generates the code in the host function that enables conditional execution of the outlined function either serially, or in redundant multithreaded mode. For certain phases of the application, a programmer/user may wish to maximize throughput performance of the system, rather than focus on the robustness of the computation. At other times, detection and correction of errors in the user-visible variable state for guaranteeing reliable execution of critical computations in an application may be more important than highest throughput performance. The adaptive RMT execution of the spheres of replication supports this behavior. It is based on internal control variables (ICV) that affect the application program behavior but are manipulated by a runtime system. Figure \ref{Fig:Code-Outlining-Adaptive} shows a code transformation which introduces optional RMT execution for the outlined code block. 

The ICVs are given values at various times during the execution of the program, and therefore their values control when the runtime enables/disables the redundant multithreading-based error detection/correction. The ICVs also control the number of redundant threads created. The ICVs are initialized by the runtime, and the application is signaled when their value is modified. When the execution encounters the outlined sphere of replication, whether the execution is performed sequentially or whether the code block is bound to dual threads (for error detection) or triple redundant threads (for correction via majority voting) depends on the current value of the ICV.

\section{The RedThreads Runtime System}
\label{sec:Runtime}

The role of the RedThreads runtime system is to support the management of the redundant threads and to provide an environment for the dynamic runtime optimizations. RedThreads provides a runtime library (RTL) and a set of accompanying environment variables (\ref{subsec:RTL}). The runtime also implements policies for the opportunistic use of RMT during application execution (\ref{subsec:opportunistic}).

\subsection{The RedThreads Runtime Library (RTL)}
\label{subsec:RTL}
The RedThreads runtime interface is based on the idea that the compiler outlines sections of code that are to run in parallel into separate functions that can then be invoked in multiple redundant threads. The RedThreads RTL contain routines that can be used to modify the execution environment at runtime, including routines for runtime initialization and finalization, thread creation and termination manipulation of locks on memory locations. Additionally, the runtime library functions provide capabilities that enable the modulation of the redundancy level. The functions allow an application to specify the mode in which to run and to change the mode based on the needs of the user and the current state of the system. The RTL interface routines are implemented as library functions, which are visible to the compiler and linker only. Their function prototypes are contained in a header {\tt redthreads.h}. These functions are not intended to be invoked as ordinary function calls by the application developer. The complete list of library functions and their capabilities are provided in Table 1.
\begin{center}
\begin{table}
 \label{table:redthreads-lib-routines}
 \begin{tabular}{r p{7cm}}
 \hline
 \textbf{Rolex library routine} & \textbf{Capability} \\
 \hline
 {\tt \_\_redthreads\_initialize()}   & Initialization of runtime threading environment and event monitoring\\
 {\tt \_\_redthreads\_finalize()}     & Clean up of structures and termination of runtime system\\
 {\tt \_\_redthreads\_fork()}	      & Redundant thread creation: each of which executes outlined function block \\
 {\tt \_\_redthreads\_join()}	      & Synchronize redundant thread copies \\ 
 {\tt \_\_redthreads\_compare()}      & Compare output values produced by sphere of replications\\
 {\tt \_\_redthreads\_get\_strength()}& Query runtime current redundancy strength \\
 {\tt \_\_redthreads\_set\_strength()}& Set redundancy strength to select between dual, triple RMT or none \\
 {\tt \_\_redthreads\_thread\_num()}  & Get thread identifier for each redundant thread \\
 {\tt \_\_redthreads\_set\_dynamic()} & Enable RMT to be enabled/disabled opportunistically\\
 {\tt \_\_redthreads\_get\_dynamic()} & Returns {\tt .true.} if dynamic RMT policy is enabled, otherwise returns {\tt .false.}\\ 
 {\tt \_\_redthreads\_init\_lock(lock)}    & Initialize lock for shared access to data variables \\
 {\tt \_\_redthreads\_destroy\_lock(lock)} & Undefine the lock instance \\ 
 {\tt \_\_redthreads\_set\_lock(lock)}     & Forces the executing redundant thread to wait until the lock is available \\ 
 {\tt \_\_redthreads\_unset\_lock(lock)}   & Releases the executing thread from ownership of the lock \\
 {\tt \_\_redthreads\_private\_copy(variable)} & Initialize private copy of each redundant thread's variable arg \\
 \hline
 \end{tabular}
 \caption{Rolex runtime library routines}
\end{table}
\end{center}
The compiler front-end pass inserts calls to the appropriate functions based on the {\tt redthreads} directive and clauses. This shields the details of the replicated execution of the outlined functions and error checking from the programmer. The avoidance of direct threading enables portability to various low-level threading libraries and hardware architectures as well as supports better composability with other programming paradigms such as OpenMP and MPI, and gives the runtime system more freedom to allocate resources. 

\subsection{Opportunistic Fault Detection}
\label{subsec:opportunistic}
The RedThreads directives, compiler front-end and RTL routines are designed to support optional redundancy (as opposed to mandatory redundant execution, which is the norm with most redundancy-based approaches). The performance overhead of redundancy can be mitigated when the redundancy is applied opportunistically \cite{Hukerikar:2014:HPCS}. To this end the runtime monitors and analyzes patterns among fault events in the system. Various studies that seek to statistically model patterns between fault events have observed a strong correlation between error events, both spatially and temporally. 

The RedThreads runtime monitors and logs fault events in the system. The error detection and reporting features in the Advanced Configuration and Power Interface (ACPI)'s \cite{ACPI:URL} Platform Error Interface (APEI) communicate fault events to the host firmware, or to the operating system directly. Examples of such events are ECC errors on DRAM DIMMs, PCI bus parity errors, cache ECC errors, DRAM scrubbing notifications, TLB errors, memory controller errors, etc. We have developed a kernel module that captures the error logging information and communicates it to the runtime system. 
\begin{figure*} [t]
\centering
\includegraphics[width=\linewidth,height=30mm]{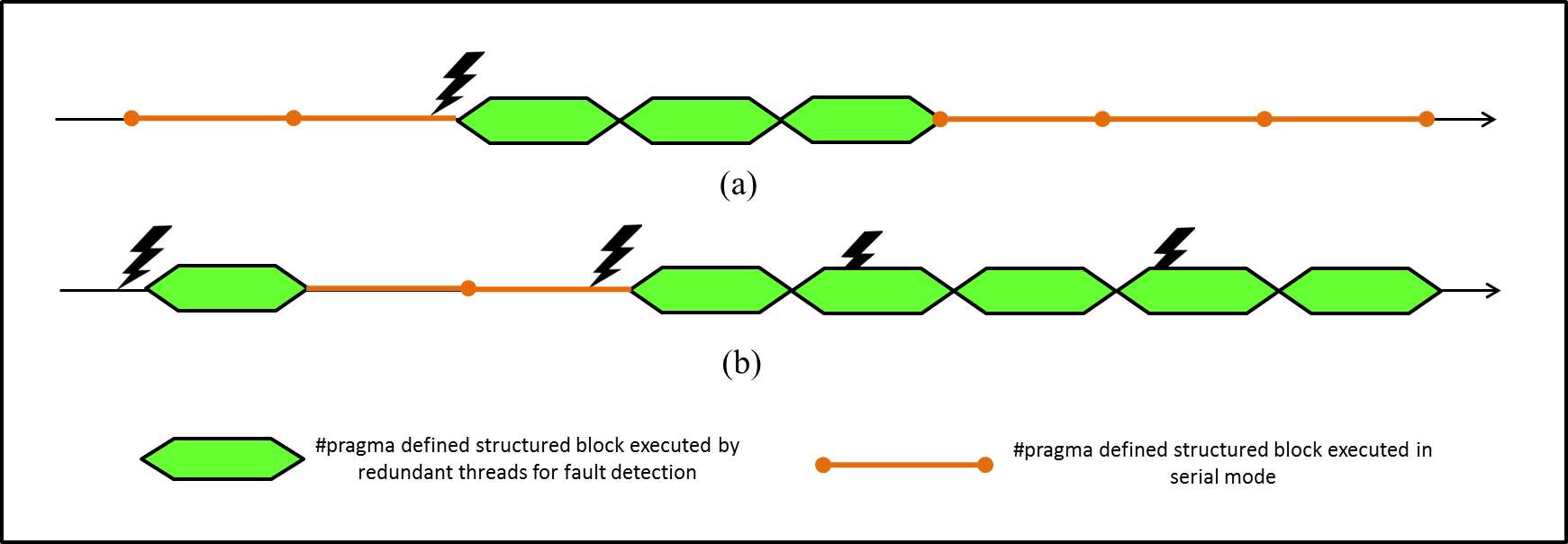}
\caption{Timeline view of adaptive redundant multithreading}
\label{fig:RMTConcept}
\end{figure*}

For the modulation of the redundancy level, the runtime seeks to model the empirical distribution of the time duration between fault events. The runtime system defines a metric called the time-between-events (TBE) and also tracks the time-since-last-event (TSLE). The aim of the algorithm is not accurate fault prediction, and therefore we do not differentiate between the types of fault events and base the adaptation algorithm on all events logged by the operating system. Since the compiler outlines every \texttt{redthreads} scoped code region to execute optionally in serial or RMT mode (Figure \ref{Fig:Code-Outlining}), the application execution always initializes in serial mode. Upon the occurrence of the first fault event, the runtime modifies an ICV to enable the RMT mode. The subsequent spheres of replication are executed in RMT. When the runtime observes no further fault events for an interval longer than average TBE, i.e., TSLE \textgreater TBE, it resets the ICV and passes a signal to the application to disable the RMT mode. When the fault event frequency increases, the runtime once again enables the RMT mode. The Figure \ref{fig:RMTConcept} illustrates the timeline view of the execution of such an application code. For the scenario in Figure \ref{fig:RMTConcept}(a), the runtime enables the RMT mode after the fault event and subsequently disables it. The application run in Figure \ref{fig:RMTConcept}(b) experiences a burst of fault events, which causes the runtime to enable RMT execution for the remainder of the application execution. From an application's perspective, the opportunistic use of RMT enables the error detection/correction capabilities when there are frequent fault events in the system and the likelihood of errors in the program state is high. During periods of relative stability, in which no fault events are logged, the detection/correction capabilities are disabled in order limit the performance overhead of using RMT.

\section{Runtime Optimization Strategies for RedThreads}
\label{sec:Optimization}

The key benefit of a software-based RMT approach is the flexibility in scheduling the redundant threads by the runtime system. We explore two optimization strategies to the opportunistic RMT to seek to further minimize the performance overhead of the redundancy on the application's execution: a lazy fault detection scheme and a thread clustering scheme.   

\subsection{Lazy Fault Evaluation}
In the adaptive RMT approach, there is an implicit barrier at the end of a RedThreads scoped code block. When the block is executed by redundant threads copies, the threads must synchronize and compare values produced by each thread before the execution proceeds. The amount of computation that is enclosed within the programmer-defined sphere of replication is highly application dependent. For applications in which long phases are enclosed in the pragma scoped block, or in compute-bound applications, the requirement that the output values be compared after each instance of the structured blocks places an unnecessary constraint that incurs a nontrivial overhead on the overall application performance.

\begin{figure*} [t]
\centering
\includegraphics[width=\linewidth,height=30mm]{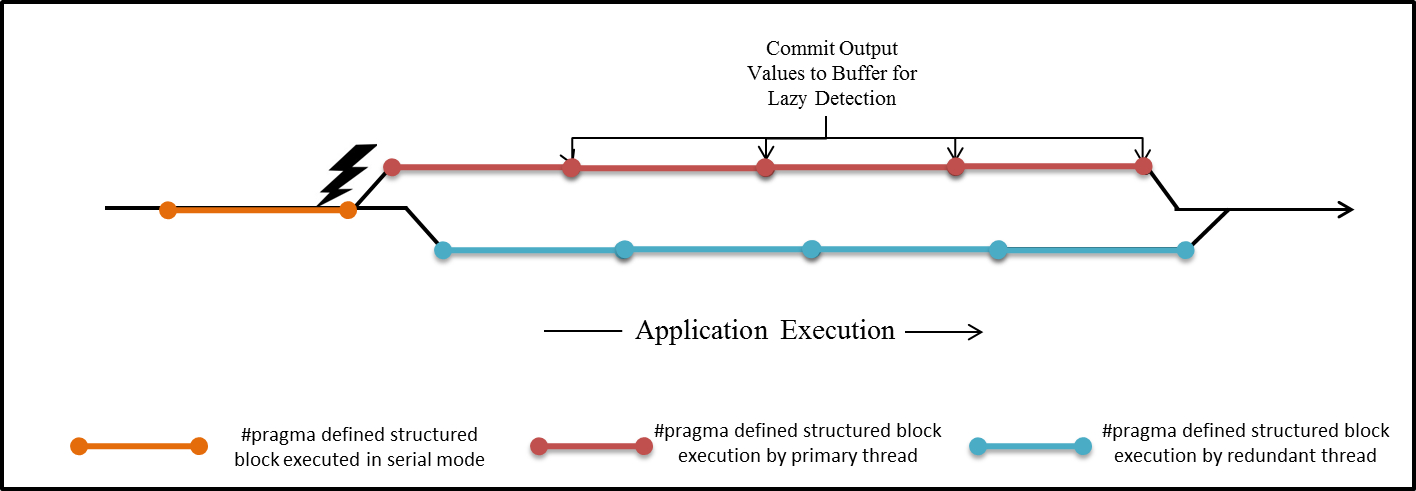}
\caption{Timeline view of adaptive redundant multithreading with lazy evaluation}
\label{fig:LazyRMTConcept}
\end{figure*}

To address this issue, we developed a \textit{lazy} fault detection scheme \cite{Hukerikar:2014:HPEC}, in which we relax the requirement that the redundant threads synchronize immediately after the code block. The \textit{lazy} strategy is based on the insight that the error detection need not lie on the application's critical path. The redundant threads can execute subsequent code blocks independently, without the need to synchronize the threads and perform output value comparison. The runtime system allocates buffer space into which the duplicate threads can write their respective output values as they execute each block. A separate lightweight fault detection thread performs the value comparison. Figure \ref{fig:LazyRMTConcept} illustrates the lazy evaluation approach. 

The output buffer is implemented as a cyclic FIFO. Each entry of the buffer contains a pair of elements and the size of each buffer entry is determined based on the data types of the variable list specified in the {\tt compare} scoping clause. Every circular buffer entry is protected by a lock since it is accessed by redundant threads as well as the detection thread. For each individual element in the buffer entry there is a flag to set its validity. When the detection thread is scheduled, it performs a value comparison for all buffer entries for which both elements in an entry pair are in valid state. Each buffer entry contains an \textit{active} flag that tracks whether either element in the buffer entry is currently in use. This enables the runtime to reclaim buffer space when the value comparison is complete.

\subsection{Thread Clustering}
In our adaptive RMT approach, the redundant threads are assigned to the available processor cores such that the application workload is evenly distributed. In order to improve the overall application performance, we explore a thread scheduling solution that assigns variable priorities to the redundant threads. An elevated priority is assigned to one of the redundant threads for each outlined sphere of replication. The runtime assigns a lower scheduling priority to the duplicate threads. Additionally, we statically assign the CPU affinity of the threads such that all the duplicate threads are clustered together and assigned to a single processor core. The primary threads for each structured block are balanced across the remaining processor cores. The thread clustering yields an unbalanced thread-to-core mapping in which a single processor core running the redundant threads tends to be oversubscribed while the primary threads contend with one fewer processor core. The intent of this strategy is to seek to reduce the interference between the primary and redundant threads, which amortizes some the overhead incurred by the redundant computation. 

When combined with the lazy evaluation approach, the relaxation of the implicit synchronization barrier for every code block enables varied scheduling priorities and clustering of the primary and redundant threads to separate processor cores. This permits a primary thread to proceed with the application execution, without stalling on the redundant threads to complete their execution of the same code blocks. The dedicated output value comparison thread is also assigned a lower scheduling priority than the primary thread. The runtime system sets the CPU affinity of the primary, redundant and detection threads based on the static assignments expressed through a RTL routine. For systems containing large number of cores, the RTL enables the creation of \textit{islands}, i.e., groups of processor cores in which a single processor core in the group is assigned the redundant computation, while the remaining cores execute the primary threads.

\section{Experimental Evaluation}
\label{sec:Experimental_Evaluation}

\subsection{Scoping Spheres of Replication using RedThreads}
We study the performance and fault coverage impact of applying RedThreads for error detection \& correction in various scientific application codes. For each code we identify and scope the most critical computational phases to create spheres of replication: 

\textbf{Double-Precision Matrix-Matrix Multiplication (DGEMM)}
For the DGEMM code, we define the scope of the RedThreads pragma block to include the inner dot product of the matrix multiplication, \textit{i.e.,} the dot product computation resulting from the multiplication of a single row and single column of the operand matrices.

\textbf{Sparse Matrix Vector Multiplication (SpMV)}:
For the SpMV code, we scope the inner product loop in the {\tt \#pragma redthreads} block. The reduced vector element y[i] for every row is the output value that is compared to detect the presence of errors for each iteration. 

\textbf{Conjugate Gradient (CG)}
The conjugate gradient (CG) code solves a system of linear equations through the iterative refinement of an initial approximate solution. The iterations provide a monotonically decreasing residual error, and by enclosing each iteration in the {\tt redthreads} block the residual norm is compared to detect the presence of errors in each CG iteration. 

\textbf{Self-Stabilizing Conjugate Gradient (SS-CG)}
The self-stabilizing approach to the conjugate gradient method \cite{Sao:2013} contains a correction step that ensures that the algorithm remains in valid state. We only include the code for the correction step in the {\tt \#pragma redthreads} structured code block. No fault detection/correction is necessary for the remaining CG iterations since the correction step accounts for any computational errors in those iterations and restores the stability of the algorithm.

\textbf{Multigrid Solver}
The algebraic multigrid method solves a coarse approximation of the original problem, interpolates it back to a finer level and then refines that solution until it forms a sufficiently precise solution. We use a V-cycle depth of 8, and the {\tt \#pragma redthreads} directive is used to provide fault detection coverage for the code blocks corresponding to the restriction, relaxation and interpolation phases of the V-cycle.

\subsection{Evaluation Methodology}
In order to experimentally evaluate the adaptive redundancy offered by RedThreads, we perform a series of fault injection experiments. Each application code described above is implemented in C and parallelized to use multithreading through OpenMP directives. For each application code, we embed {\tt \#pragma redthreads} directive in place of the OpenMP directive to define the scope of the sphere of replication. The codes are transformed using our ROSE-based front-end, compiled by a GCC backend, and linked with the RedThreads library. The experiments are performed on a Dell PowerEdge 1950 2.5GHz 4-core/dual socket compute node in the USC HPCC cluster \cite{USC:HPCC}. The node uses the Intel\textregistered Xeon\textregistered 5300 series processors and contains 12GB DRAM memory. The node operating system is CentOS Linux v6.5. 

In the first set of experiments, we analyze the fault coverage provided by the adaptive RMT. The fault injection framework is able to inject multiple random faults to a single instance of program execution. To measure the sensitivity of the application to transient errors, we use an application robustness rating, which is the inverse of the error rate for which the application experiences failure. A higher rating indicates higher application resilience. We run fault injections to compare the baseline (without adaptive RMT) rating with the effective robustness rating with adaptive RMT support. To achieve statistical confidence for the results, each application code is run 10,000 times and the interval between any two consecutive faults is randomized for each application run. 

We also evaluate the performance overhead of the adaptive RMT strategy. In these performance analysis experiments, the fault injection framework generates error notification events. These simulated events are notifications, which do not perturb any aspect of the application program state. These events are logged by the runtime and used by the RMT adaptation algorithm. The fault injection framework generates events by sending USR1 signals to an application process. The runtime library, which is linked to the application process, contains a handler for these signals. The signal handler is a blocking routine which catches and logs the fault event notification. We use a large number of execution runs (10,000 per application code) with randomized fault intervals, which allows us to observe the average application overhead for a variety of fault patterns for each injection rate.

\subsection{Results}
In order to evaluate the performance viability of applying the language directive to scope the spheres of replication for each code, we compare the RedThreads-based adaptive RMT with the most widely used redundancy approach in HPC applications, i.e., process-level replication. For this analysis we do not enable the adaptation algorithm, which means all RedThreads scoped regions are executed in RMT mode. We compare the performance of the RedThreads scoped RMT with a serial implementation, and with an implementation that uses process-level duplication for each code. The average normalized execution time for the process replication incurs an overhead in the excess of 100\% over the serial implementations. However, for codes such as the SS-CG, in which the algorithm-based fault resilience features are combined with RMT, the time-to-solution with RMT is only 1.4x of a serial implementation, rather than 2.1x with process replication. For the other codes, in which a significant portion of the computation is executed using RMT, the fact that the redundant threads do not replicate all the computations, and the common address space and shared access to the operand data, enables at least 10 to 15\% lower overhead than process-level replication.
\begin{figure}[tp]
\centering
\includegraphics[width=\linewidth,height=65mm]{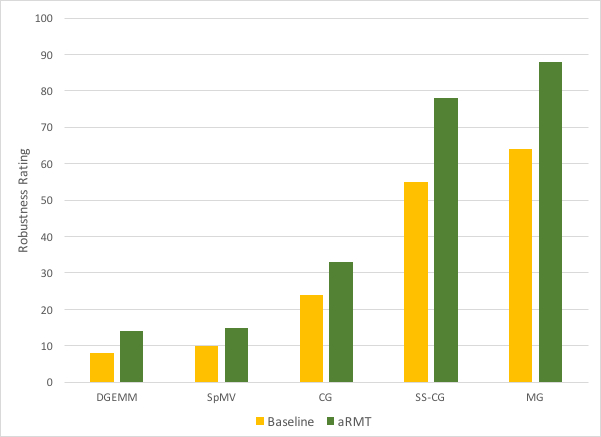}
\caption{Results: Fault Coverage Analysis based on Application Robustness Rating}
\label{Fig:Results-VF}
\end{figure}

The analysis of the robustness ratings of each code reveals different sensitivities to errors (Figure \ref{Fig:Results-VF}). The multigrid solver, the CG and SS-CG solvers are based on iterative algorithms, which refine an initial solution until its error bound is below a certain threshold. Their intermediate solution state is usually tolerant to perturbations and any errors in such state are often not fatal. Providing fault coverage to the solver iterations using RMT improves the robustness rating, and the application resilience for each of these solver codes. The DGEMM and SpMV codes demonstrate higher sensitivity to errors in any aspects of their program states. For these experiments, in order to calculate the application's robustness rating, the faults are injected at accelerated rates. For these rates, only modest improvements in robustness rating are observed. For lower rates, the runtime modulates the fault coverage based on observed fault rates and the improvement in robustness rating is substantially higher.    

\begin{figure}
\centering
\includegraphics[height=180mm,width=\linewidth]{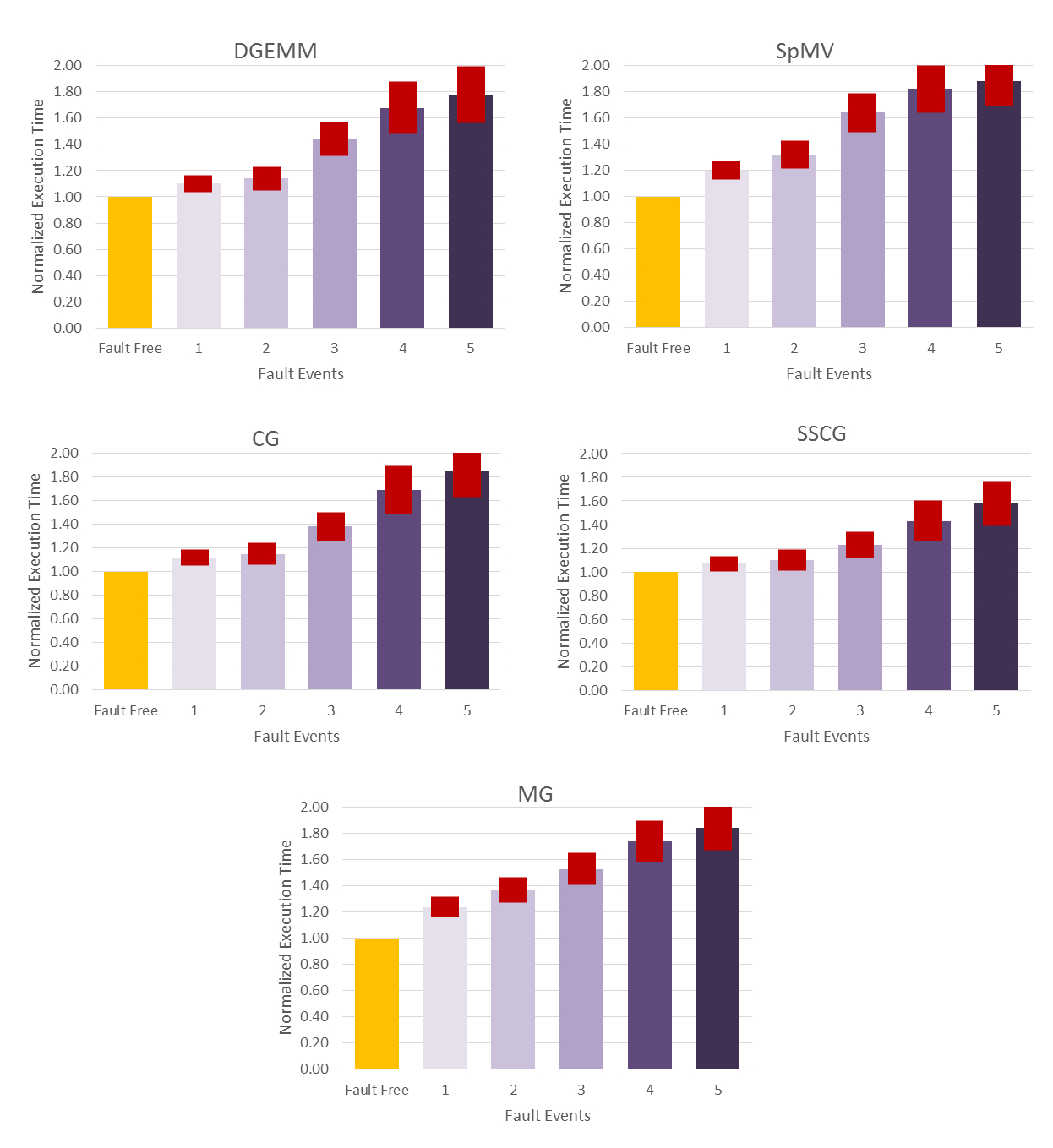}
\caption{Results: Fault detection with adaptive redundant multithreading}
\label{Fig:Results-Fault-Detection}
\end{figure}

\begin{figure}
\centering
\includegraphics[height=180mm,width=\linewidth]{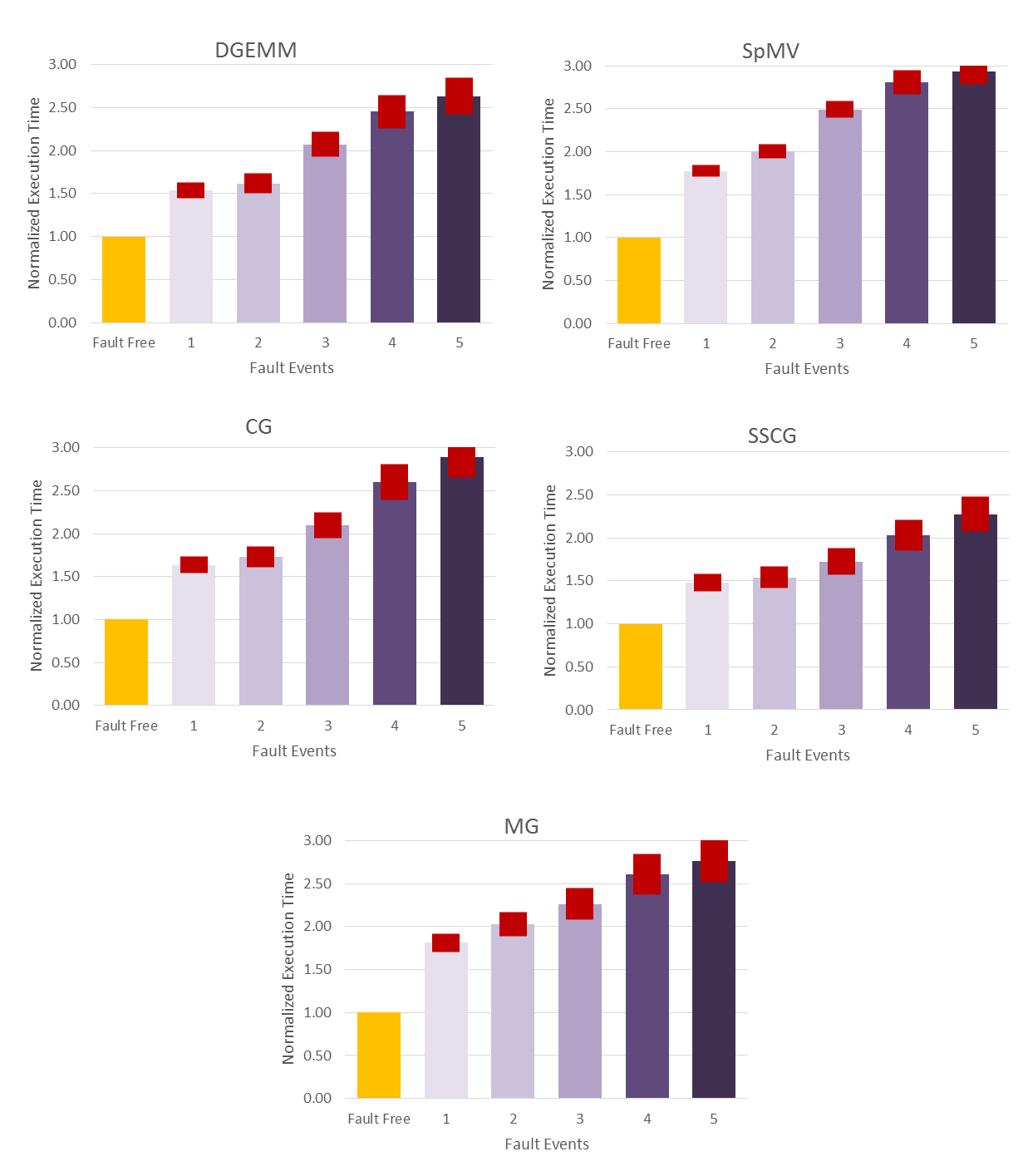}
\caption{Results: Fault detection and correction with adaptive redundant multithreading}
\label{Fig:Results-Fault-Correction}
\end{figure}

Figure \ref{Fig:Results-Fault-Detection} shows a summary of the results of the performance evaluation experiments with fault injections for which the adaptive RMT is enabled. The strength of RMT is set to provide fault detection using DMR. Each data point is the average normalized execution time for the 10,000 application runs. The error bars illustrate the spread among the maximum and minimum overhead for each fault injection rate. For the fault injection rate of 1 event per execution run, the adaptive RMT strategy incurs 7\% to 24\% overhead over a fault-free execution. For the higher fault injection rates, the spread among the execution times is higher since the experiments cover a range of fault patterns. However, the maximum average overhead incurred is in the range of 78\% to 85\% which is still lower than an overhead in the excess of 100\% that is typical for full-process replication. The key insight offered by these results is that the performance overhead tracks the injected rate. Therefore, the adaptive RMT effectively supports dynamic tuning of the fault protection and its performance overhead. We also study the performance impact of applying TMR through adaptive RMT.
Figure \ref{Fig:Results-Fault-Correction} shows the results of the TMR performance evaluation with fault injections. For the low fault rates, the overhead to the time to solution is in the range of 40\% to 74\%. For a fault rate of up to 3 faults per execution run, the overhead incurred for triple adaptive RMT is lower than or at least proportional to the overhead due to dual redundant full-process replication, for most of the application codes we evaluated. For higher fault rates, the overheads while less than the 3x overheads typical to full-process TMR, are substantial (in the range of 2.5x to 2.9x). These results suggest that triple RMT-based detection may be a viable strategy for detection and correction at low fault rates. In the presence of higher fault rates, the use of dual RMT-based error detection in collaboration with a different recovery, or amelioration strategy, might be a more suitable resilience solution for long-running computations.

\begin{figure}
\centering
\includegraphics[height=180mm,width=\linewidth]{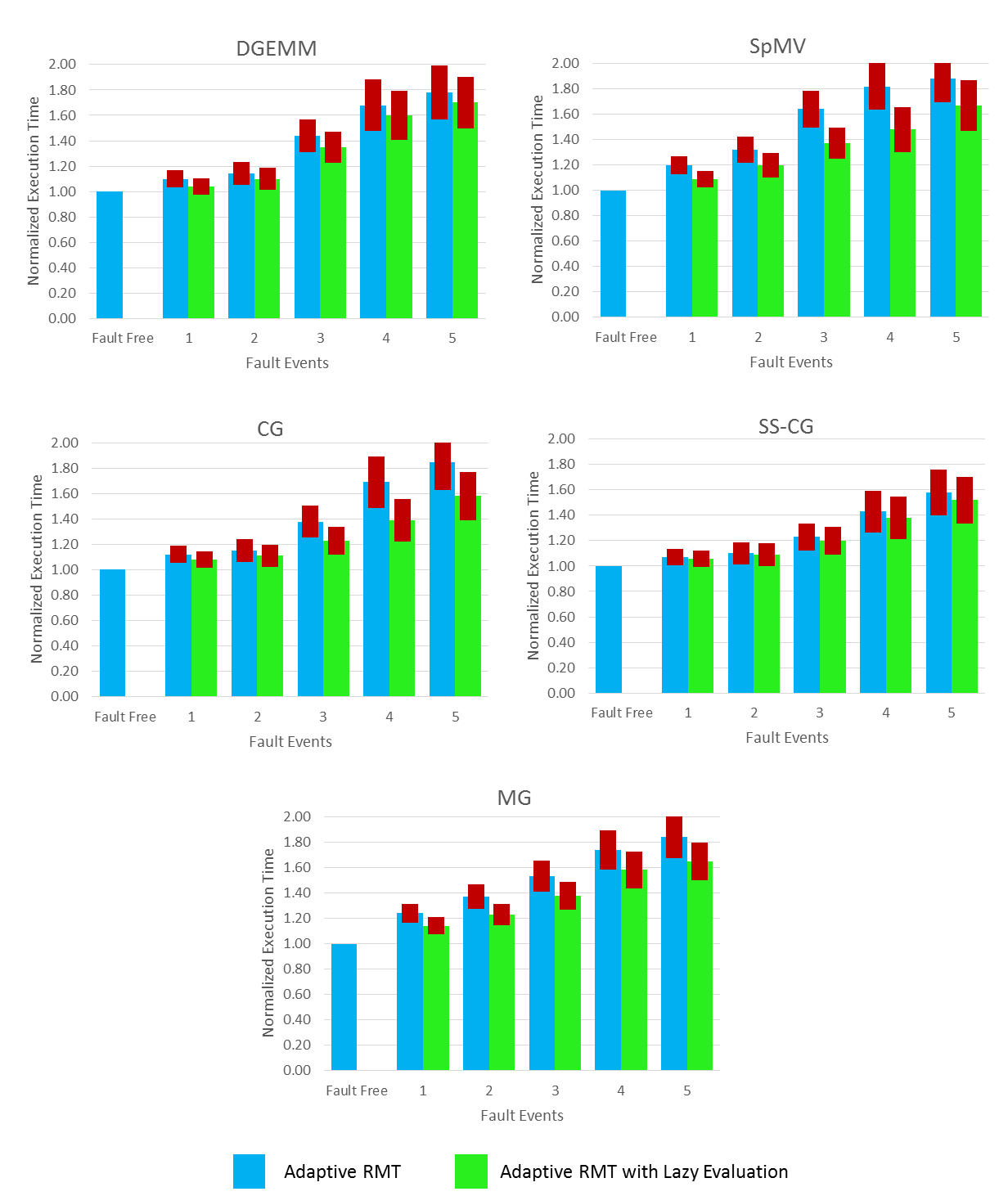}
\caption{Results: Fault detection with adaptive RMT with lazy evaluation}
\label{Fig:Results-Fault-Detection-Lazy}
\end{figure}

\begin{figure}
\centering
\includegraphics[height=180mm,width=\linewidth]{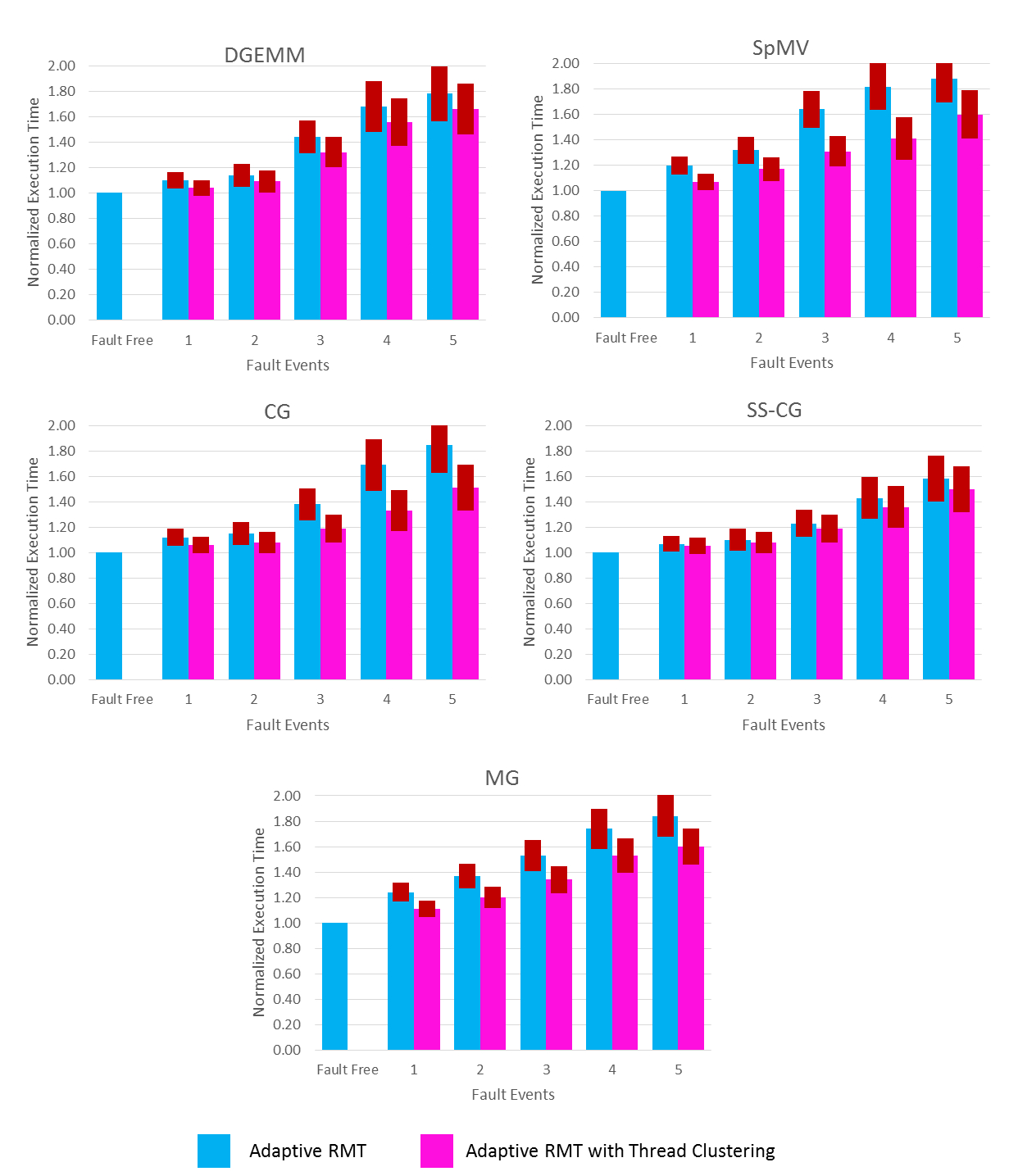}
\caption{Results: Fault detection with adaptive RMT with thread clustering}
\label{Fig:Results-Fault-Detection-TC}
\end{figure}

The results in Figure \ref{Fig:Results-Fault-Detection-Lazy} summarize the performance overhead analysis for optimization based on lazy detection. Here, relaxation of the requirement that the redundant threads synchronize and compare values after every instance of the structured code block allows the primary thread to race ahead while the redundant thread lags behind. In comparison to the time to solutions for the codes in Figure \ref{Fig:Results-Fault-Detection}, the lazy evaluation provides a further 10 to 18\% improvement in the time-to-solution. Figure \ref{Fig:Results-Fault-Detection-TC} presents the results of the experiments that use the runtime scheduling optimizations based on thread clustering. These results show that assigning a higher priority value and dedicating core resources to the primary computation yields a better time-to-solution for each of the application codes. The use of unbalanced scheduling yields higher performance gains at higher fault rates when a larger portion of total computation is executed in RMT mode.

\section{Related Work}
\label{sec:Related_Work}

\subsection{Hardware-based Redundant Execution Schemes}
In contrast to error-coding techniques \cite{Moon:2005} that detect and correct faults using redundant information in storage bits or logic units, the use of redundant execution enables higher fault coverage since it typically protects the logic and computation blocks as well as the data these operate on. Redundant execution techniques are based on either lockstepping, or redundant multithreading. Lockstepping requires comparison of the sphere of replication at every clock cycle. Lockstepping implementations used in the design of highly reliable server architectures include the Stratus ftServer \cite{Somers:2002}, the Hewlett-Packard NonStop Architecture \cite{Bernick:2005}, and the IBM Z-series processors \cite{Slegel:1999}. The need for cycle-synchronized redundant hardware often requires physically redundant identical copies of the hardware.   

The use of redundant multithreading for error detection is implemented using hardware-based threads. The early hardware-based RMT approaches concurrently executed multiple explicit copies of the program code by sharing the processor resources among the threads for error detection. These redundant execution was transparent to the system software and was proposed to mitigate the costs of implementing complete hardware replication used by lock stepped processors such as the IBM G5 \cite{Slegel:1999}. Later approaches were based on simultaneous multithreading (SMT) \cite{Vijaykumar:2002}, often on a single processor core and used the flexibility of instruction scheduling offered by the microarchitectures in superscalar processors. As chip multiprocessors (CMPs) became increasingly ubiquitous, there were studies \cite{Mukherjee:2002} \cite{Reinhardt:2000} that demonstrated the viability of executing the redundant threads across separate processor cores. Partial RMT approaches such as SliCK \cite{Parashar:2006} mitigate the contention for the core's memory and execution resources amongst the redundant threads by applying RMT only on specific instruction slices. The dynamic implementation verification architecture (DIVA) \cite{Austin:1999} uses a trailing thread implementation of RMT, in which the redundant execution is performed by an in-order pipelined checker core, while the primary instructions execution is performed on an out-of-order processor. In general, with the use of hardware-based redundant execution streams, every instruction execution must be replicated. For any program execution this multiplies the total number of instructions by at least a factor of the degree of replication.  

\subsection{Software-based Redundancy Schemes} 
Software-based redundancy approaches support error detection/correction based on compiler analysis and transformation duplicate the program code either through operating system visible redundant processes \cite{Shye:2009}, or through threads within the same process context. This enables more flexibility in terms of selection of instructions to replicate and in the assignment of redundant execution streams to hardware resources \cite{Wang:2007}. The DAFT \cite{Zhang:2010} approach uses a compiler transformation to duplicate an entire program in a redundant thread that trails the main thread and inserts instructions for error checking. The SWIFT \cite{Reis:2005} and EDDI \cite{Oh:2002} duplicate all instructions and insert ``compare" instructions to validate the program correctness at appropriate locations in the program code. The ROSE::FTTransform \cite{Lidman:2012} applies source-to-source translation to duplicate individual source-level statements to detect transient processor faults. Despite the flexibility of software-based approaches in defining certain aspects of the sphere of replication, the need for total replication of the program code incurs significant (at least 2x) overhead to application performance. In contrast, our RedThreads interface enables a more precise definition of the scope of the sphere of replication. Additionally, the RedThreads runtime system does not enforce replicated execution for the total duration of an application program execution; rather it takes a more nuanced approach in applying the redundancy through adaptive and dynamic enablement/disablement of the RMT. 

\subsection{Hardware-Software Co-design Techniques for Fault Tolerance}
Cross-layer resilience techniques \cite{Cheng:2016} employ multiple error resilience techniques from different layers of the system stack to collaboratively achieve error resilience. These frameworks combine selective circuit-level hardening and logic-level parity checking with algorithm-based fault tolerance methods to provide resilient operation. In the context of redundant execution, an adaptive scheme for RMT \cite{Gomaa:2005} proposed turning off the trailing thread, thereby reducing fault coverage and increasing fault vulnerability, during regions of a program that have high instructions per cycle (IPC). The use of chip temperature data has also been used to disable redundant thread in response to elevated operating temperatures \cite{Siddiqua:2009}. The real-time computation of the architectural vulnerability factor (AVF) based on thermal changes in a chip has been used to modulate the use of dual modular redundancy \cite{Vadlamani:2010}. In contrast, our approach uses RAS events to analyze the event intervals at the runtime system-level to enable/disable RMT.

\section{Conclusion}
\label{sec:Conclusion}

As faults will become increasingly prevalent in future exascale-class HPC systems, maintaining error resilience during the execution of long-running scientific applications is critical. Techniques that enable error detection and correction without burdening the performance will be important for these applications. In this paper, we presented a method for application-level fault detection and correction based on redundant multithreading. The RedThreads API provides language extensions to C/C++ for scoping spheres of replication and the associated variable state at the application source code level. The compiler framework through source-level transformations sets up the code regions to be optionally executed using RMT. In addition to an API to capture programmer knowledge on scoping the spheres of replication, the RedThreads solution stack contains a runtime system that continuously learns about the fault-tolerance state of the system using error notifications and modulates the redundancy based on the threat to the application program state. Our experiments demonstrated that the adaptive RMT offers coverage to the critical computations and the resulting overhead to application performance using RMT opportunistically is substantially better than na\"{i}ve, macroscale redundancy strategies.


\bibliographystyle{spmpsci}      
\bibliography{main}              

\begin{thebibliography}{10}
\providecommand{\url}[1]{{#1}}
\providecommand{\urlprefix}{URL }
\expandafter\ifx\csname urlstyle\endcsname\relax
  \providecommand{\doi}[1]{DOI~\discretionary{}{}{}#1}\else
  \providecommand{\doi}{DOI~\discretionary{}{}{}\begingroup
  \urlstyle{rm}\Url}\fi

\bibitem{ASASC:2010}
The {O}pportunities and {C}hallenges of {E}xascale {C}omputing.
\newblock Tech. rep., Summary {R}eport of the {A}dvanced {S}cientific
  {C}omputing {A}dvisory {C}ommittee ({A}{S}{C}{A}{C}) {S}ubcommittee (2010)

\bibitem{ACPI:URL}
Advanced configuration and power interface ({A}{C}{P}{I}).
\newblock \url{http://www.uefi.org/acpi/specs} (2013)

\bibitem{Austin:1999}
Austin, T.M.: Diva: A reliable substrate for deep submicron microarchitecture
  design.
\newblock In: Proceedings of the 32Nd Annual ACM/IEEE International Symposium
  on Microarchitecture, pp. 196--207 (1999)

\bibitem{Bernick:2005}
Bernick, D., Bruckert, B., Vigna, P.D., Garcia, D., Jardine, R., Klecka, J.,
  Smullen, J.: Nonstop\textregistered advanced architecture.
\newblock In: Proceedings of the 2005 International Conference on Dependable
  Systems and Networks, DSN '05, pp. 12--21 (2005)

\bibitem{Borkar:2005:Micro}
Borkar, S.: Designing {R}eliable {S}ystems from {U}nreliable {C}omponents:
  {T}he {C}hallenges of {T}ransistor {V}ariability and {D}egradation.
\newblock IEEE Micro \textbf{25}(6), 10--16 (2005)

\bibitem{Cheng:2016}
Cheng, E., Mirkhani, S., Szafaryn, L.G., Cher, C.Y., Cho, H., Skadron, K.,
  Stan, M.R., Lilja, K., Abraham, J.A., Bose, P., Mitra, S.: Clear: Cross-layer
  exploration for architecting resilience - combining hardware and software
  techniques to tolerate soft errors in processor cores.
\newblock In: Proceedings of the 53rd Annual Design Automation Conference, DAC
  '16, pp. 68:1--68:6 (2016)

\bibitem{Dongarra:2011:IES}
Dongarra, J., Beckman, P., Moore, T., et~al.: The international exascale
  software project roadmap.
\newblock International Journal on High Performance Computing Applications pp.
  3--60 (2011)

\bibitem{DARPA_ExascaleResilienceStudyReport:2009}
Elnozahy, E., Bianchini, R., El-Ghazawi, T., et~al.: System {R}esilience at
  {E}xtreme {S}cale, {W}hite {P}aper.
\newblock Tech. rep., DARPA (2009)

\bibitem{Engelmann:2009}
Engelmann, C., Ong, H.H., Scott, S.L.: The case for modular redundancy in
  large-scale high performance computing systems.
\newblock In: Proceedings of the 27th IASTED International Conference on
  Parallel and Distributed Computing and Networks (PDCN), pp. 189--194 (2009)

\bibitem{Ferreira:2011}
Ferreira, K., Stearley, J., Laros III, J.H., et~al.: Evaluating the viability
  of process replication reliability for exascale systems.
\newblock In: Proceedings of 2011 International Conference for High Performance
  Computing, Networking, Storage and Analysis, pp. 1--12 (2011)

\bibitem{Gomaa:2005}
Gomaa, M.A., Vijaykumar, T.N.: Opportunistic transient-fault detection.
\newblock SIGARCH Computer Architecture News pp. 172--183 (2005)

\bibitem{Hoemmen:2011}
Hoemmen, M., Heroux, M.A.: Fault-tolerant iterative methods via selective
  reliability.
\newblock In: Proceedings of the 2011 International Conference for High
  Performance Computing, Networking, Storage and Analysis (SC). IEEE Computer
  Society, vol.~3, p.~9 (2011)

\bibitem{Hukerikar:2014:HPCS}
Hukerikar, S., Diniz, P.C., Lucas, R.F., Teranishi, K.: Opportunistic
  application-level fault detection through adaptive redundant multithreading.
\newblock In: International Conference on High Performance Computing Simulation
  (HPCS), pp. 243--250 (2014).
\newblock \doi{10.1109/HPCSim.2014.6903692}

\bibitem{Hukerikar:Rolex:2016}
Hukerikar, S., Lucas, R.F.: Rolex: Resilience-oriented language extensions for
  extreme-scale systems.
\newblock The Journal of Supercomputing pp. 1--33 (2016).
\newblock \doi{10.1007/s11227-016-1752-5}

\bibitem{Hukerikar:2014:HPEC}
Hukerikar, S., Teranishi, K., Diniz, P.C., Lucas, R.F.: An evaluation of lazy
  fault detection based on adaptive redundant multithreading.
\newblock In: IEEE High Performance Extreme Computing Conference (HPEC), pp.
  1--6 (2014).
\newblock \doi{10.1109/HPEC.2014.7040999}

\bibitem{DARPA_ExascaleTechStudyReport:2008}
Kogge, P., Bergman, K., Borkar, S., et~al.: Exascale {C}omputing {S}tudy:
  {T}echnology {C}hallenges in {A}chieving {E}xascale systems.
\newblock Tech. rep., DARPA (2008)

\bibitem{Liao:2010}
Liao, C., Quinlan, D.J., Vuduc, R., Panas, T.: Effective source-to-source
  outlining to support whole program empirical optimization pp. 308--322 (2010)

\bibitem{Lidman:2012}
Lidman, J., Quinlan, D., Liao, C., McKee, S.: R{O}{S}{E}::{F}{T}{T}ransform - a
  {S}ource-to-{S}ource {T}ranslation {F}ramework for {E}xascale
  {F}ault-tolerance {R}esearch.
\newblock In: Dependable {S}ystems and {N}etworks {W}orkshops (DSN-W), 2012
  IEEE/IFIP 42nd International Conference on, pp. 1--6 (2012)

\bibitem{Moon:2005}
Moon, T.K.: Error correction coding: Mathematical methods and algorithms
  (2005)

\bibitem{Mukherjee:2002}
Mukherjee, S.S., Kontz, M., Reinhardt, S.K.: Detailed {D}esign and {E}valuation
  of {R}edundant {M}ultithreading {A}lternatives.
\newblock SIGARCH Computer Architecture News pp. 99--110 (2002)

\bibitem{vonNeumann:1956}
von Neumann, J.: Probabilistic {L}ogics and the {S}ynthesis of {R}eliable
  {O}rganisms from {U}nreliable {C}omponents.
\newblock Automata Studies pp. 43--98 (1956)

\bibitem{Oh:2002}
Oh, N., Shirvani, P.P., McCluskey, E.J.: Error {D}etection by {D}uplicated
  {I}nstructions in {S}uper-scalar {P}rocessors.
\newblock IEEE {T}ransactions on {R}eliability pp. 63--75 (2002)

\bibitem{Parashar:2006}
Parashar, A., Sivasubramaniam, A., Gurumurthi, S.: Slick: Slice-based locality
  exploitation for efficient redundant multithreading.
\newblock SIGOPS Operating Systems Review (5), 95--105 (2006)

\bibitem{ROSE:Compiler}
Quinlan, D., et~al.: Rose {C}ompiler (2000).
\newblock \urlprefix\url{http://www.rosecompiler.org}

\bibitem{Reinhardt:2000}
Reinhardt, S.K., Mukherjee, S.S.: Transient {F}ault {D}etection via
  {S}imultaneous {M}ultithreading.
\newblock In: Proceedings of the 27th Annual International Symposium on
  Computer Architecture, pp. 25--36 (2000)

\bibitem{Reis:2005}
Reis, G., Chang, J., Vachharajani, N., et~al.: S{W}{I}{F}{T}: {S}oftware
  {I}mplemented {F}ault {T}olerance.
\newblock In: International Symposium on Code Generation and Optimization,
  2005, pp. 243--254 (2005)

\bibitem{Sao:2013}
Sao, P., Vuduc, R.: Self-stabilizing iterative solvers.
\newblock In: Proceedings of the Workshop on Latest Advances in Scalable
  Algorithms for Large-Scale Systems, ScalA '13, pp. 4:1--4:8 (2013)

\bibitem{Shye:2009}
Shye, A., Blomstedt, J., Moseley, T., Reddi, V.J., Connors, D.A.: Plr: A
  software approach to transient fault tolerance for multicore architectures.
\newblock IEEE Transactions on Dependable and Secure Computing \textbf{6}(2),
  135--148 (2009)

\bibitem{Siddiqua:2009}
Siddiqua, T., Gurumurthi, S.: Balancing soft error coverage with lifetime
  reliability in redundantly multithreaded processors.
\newblock In: 2009 IEEE International Symposium on Modeling, Analysis
  Simulation of Computer and Telecommunication Systems, pp. 1--12 (2009)

\bibitem{Slegel:1999}
Slegel, T., Averill R.M., I., Check, M., et. al: I{B}{M}'s {S}/390 {G}5
  {M}icroprocessor {D}esign.
\newblock Micro, IEEE pp. 12--23 (1999)

\bibitem{Somers:2002}
Somers, J.: Stratus ftserver--intel fault tolerant platform.
\newblock Intel Developer Forum  (2002)

\bibitem{Stearley:2012}
Stearley, J., Ferreira, K., Robinson, D., et~al.: Does {P}artial {R}eplication
  {P}ay off?
\newblock In: IEEE/IFIP 42nd International Conference on Dependable Systems and
  Networks Workshops (DSN-W) (2012)

\bibitem{USC:HPCC}
USC: Center for high-performance computing.
\newblock \urlprefix\url{https://hpcc.usc.edu/}

\bibitem{Vadlamani:2010}
Vadlamani, R., Zhao, J., Burleson, W., Tessier, R.: Multicore soft error rate
  stabilization using adaptive dual modular redundancy.
\newblock In: Proceedings of the Conference on Design, Automation and Test in
  Europe, DATE '10, pp. 27--32 (2010)

\bibitem{Vijaykumar:2002}
Vijaykumar, T., Pomeranz, I., Cheng, K.: Transient-{F}ault {R}ecovery using
  {S}imultaneous {M}ultithreading.
\newblock In: 29th Annual International Symposium on Computer Architecture,
  2002, pp. 87--98 (2002)

\bibitem{Wang:2007}
Wang, C., Kim, H., Wu, Y., Ying, V.: Compiler-{M}anaged {S}oftware-based
  {R}edundant {M}ulti-{T}hreading for {T}ransient {F}ault {D}etection.
\newblock In: International Symposium on Code Generation and Optimization,
  2007, pp. 244--258 (2007).
\newblock \doi{10.1109/CGO.2007.7}

\bibitem{Zhang:2010}
Zhang, Y., Lee, J.W., Johnson, N.P., August, D.I.: D{A}{F}{T}: {D}ecoupled
  {A}cyclic {F}ault {T}olerance.
\newblock In: Proceedings of the 19th international conference on Parallel
  architectures and compilation techniques, PACT '10, pp. 87--98 (2010)

\end{thebibliography}

\end{document}